\documentclass[letterpaper]{jacow}
\usepackage{graphicx}
\usepackage{booktabs}
\usepackage{textcomp}
\begin{document}

\title{CONFORMAL UNCERTAINTY QUANTIFICATION OF A NORMALIZING FLOW SURROGATE FOR THE FACET-II TWO-BUNCH BEAM}

\author{R. Wu\textsuperscript{1,2}\thanks{These authors contributed equally.}, J. Dong\textsuperscript{1}\footnotemark[1], A. Edelen\textsuperscript{2}, M. J. Kochenderfer\textsuperscript{1},
        \textsuperscript{1}Stanford University, Stanford, CA, USA \\
        \textsuperscript{2}SLAC National Accelerator Laboratory, Menlo Park, CA, USA \\
        \texttt{\{rwu4, jessiedong, mykel\}@stanford.edu, edelen@slac.stanford.edu}}

\maketitle

\begin{abstract}
Fast surrogate models of the beam can be used for tuning, diagnostics, and control where end-to-end simulation is too slow. At FACET-II, a normalizing flow surrogate predicts the 6D high peak current two-bunch beam used for plasma wakefield acceleration (PWFA) in less than a second, from 37 injector and linac machine parameters (such as the RF phases controlling the bunch compression). However, a surrogate can predict an incorrect beam that seems plausible, so operators and controllers need to know when to trust its predictions. We use conformal prediction to generate a prediction-specific uncertainty with guaranteed coverage. That is, if we sample new machine parameters from the same distribution as our calibration data, the uncertainty will cover the simulation output at least 90\% of the time. We provide two implementations: prediction intervals for beam parameters with fixed and adaptive widths, and a region in the full 6D phase space covering 95\% of the electrons, 90\% of the time. The median size of this region is 5 (drive) and 2 (witness) times smaller than a Gaussian ellipsoid with the same coverage. Across 1,000 independent test simulations, our prediction intervals identify failure cases for the surrogate, specifically the slice peak current of highly compressed bunches.
\end{abstract}

\section{INTRODUCTION}

Plasma wakefield experiments at FACET-II \cite{yakimenko2019,litos2014} employ a two-bunch beam: a drive bunch excites the wake and a witness bunch is accelerated. Experiment performance depends on the 6D beam, including the inter-bunch separation and the emittance, peak current, and dispersion of each bunch. The beam must have a high peak current, which requires strong bunch compression.

End-to-end simulation of the 6D beam is too slow for tuning, diagnostics, and control, and for reinforcement learning training, which needs thousands of beams. A normalizing flow surrogate predicts the 6D beam in less than a second. However, as a fast generative model, it can predict plausible but incorrect beams, especially far from its training data \cite{scheinker2024}. Such predictions could mislead operators and controllers, so each prediction needs its own uncertainty.

Previous accelerator surrogate models have used ensembles and quantile networks for uncertainty quantification (UQ) \cite{mishra2021,convery2021}. However, these give no coverage guarantee, where coverage is the fraction of new cases in which the UQ interval covers the actual value. Conformal prediction provides this guarantee \cite{vovk2005,angelopoulos2023}.

We first describe our surrogate model and calibration data, then conformal prediction, which is still relatively new in the accelerator community. We then give prediction intervals for the beam parameters used in tuning, show where they fail, and build a region in 6D phase space that contains 95\% of the electrons.

\begin{figure*}[!tb]
  \centering
  \includegraphics*[width=0.95\textwidth]{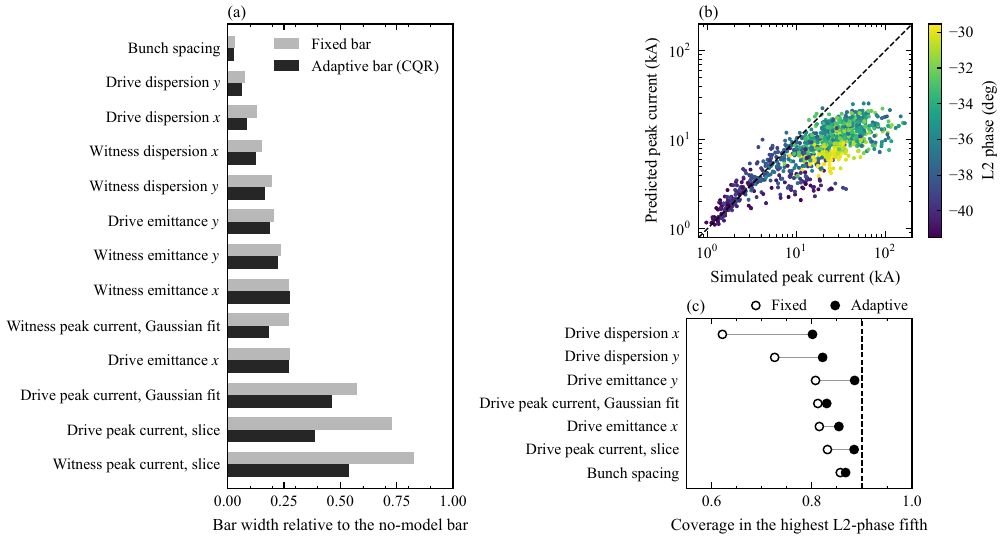}
  \caption{Conformal prediction intervals on beam characteristics, computed over 1,000 independent samples. (a) Interval width relative to the model-free interval, for fixed (gray) and adaptive (black) intervals. For emittance and peak current, the width is computed from the error in $\log y$. (b) Predicted versus simulated drive slice peak current, colored by L2 phase. The dashed line shows the ideal prediction. (c) Coverage of fixed (open circles) and adaptive (filled circles) intervals on the fifth of cases with the highest L2 phase. The dashed line shows the target coverage of 90\%.}
  \label{fig:bars}
\end{figure*}

\section{SURROGATE MODEL AND CALIBRATION DATA}

Our surrogate model is a conditional normalizing flow \cite{durkan2019} that maps a spherical Gaussian distribution to a 6D particle distribution and back. It is conditioned on 37 machine parameters that operators adjust during tuning, and the training simulations vary all 37 over the operational range: 12 injector parameters (laser timing, bunch charges, gun phase and field, solenoid, etc.) and 25 linac parameters (RF phase and energy per linac section, BC20 chicane and final focus magnets, etc.). The RF phases of L1 and L2 control the energy chirp and thus the bunch compression.

A random input, sampled once per beam, produces an ensemble of similar beams for the same parameters. We trained the flow on $\sim$11,000 FACET2-S2E simulations with 4,096 macroparticles per bunch at the transverse deflecting cavity. From the predicted particles, we compute the inter-bunch separation, x and y normalized emittance, peak current, and x and y dispersion of each bunch. The peak current is either the slice peak (maximum over 40 slices) or the peak of a Gaussian fit to the current profile. For each parameter set, we use the median over 8 sampled beams.

The calibration data are 1,000 new simulations with parameters sampled independently and uniformly from the training range. None was used for training or model selection. In 45 cases, the witness bunch is lost or largely depleted, so we use the remaining 955 for the witness beam.

For conformal prediction to work, the calibration and the new data have to be exchangeable. If the calibration and training data are too close, the intervals will be too narrow. For example, the validation set obtained during training could serve as calibration data, but 62\% of its simulations have the same injector settings as one of the training simulations, and intervals based on it cover only 80 to 91\% of new settings. If the new data comes from a different distribution, such as a control loop or the physical machine, the coverage guarantee does not apply.

\section{CONFORMAL PREDICTION}

Split conformal prediction provides a coverage guarantee for any prediction model. It requires a scoring function $s(x, y)$ that measures the prediction error of the model for some input $x$ and target $y$. After evaluating the score on $n$ calibration inputs, we set $q$ to the $\lceil (n+1)(1-\alpha) \rceil$-th smallest score, slightly above the 90th percentile for $\alpha = 0.1$. For any new input $x$, the prediction set is then $C(x) = \{y : s(x, y) \le q\}$. If the new input is exchangeable with the calibration inputs, then
\begin{equation}
  P\left[\, y_\mathrm{new} \in C(x_\mathrm{new}) \,\right] \ge 1 - \alpha .
\end{equation}
This holds regardless of the model and the error distribution. However, (1) is about the average coverage, and some regions of the input space may have lower or higher coverage. We therefore also test on subsets of the test data. The main design decision is the scoring function, as it determines the structure of the prediction sets.

\begin{figure*}[!tb]
  \centering
  \includegraphics*[width=0.95\textwidth]{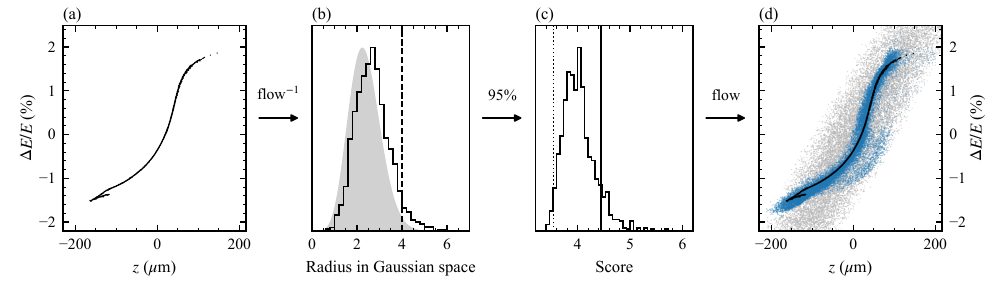}
  \caption{Construction of the 6D region for one drive bunch. (a) Simulated particles in longitudinal phase space. (b) The same particles mapped back through the flow: the distribution of their radius in the Gaussian space, against a perfect surrogate (gray). The dashed line is their 95\% radius, the score. (c) Scores of all calibration cases and the threshold $r$ (solid), against 3.55 for a perfect surrogate (dotted). (d) The ball of radius $r$ mapped through the flow (blue), the simulated particles (black), and the Gaussian ellipsoid (gray).}
  \label{fig:region}
\end{figure*}

\section{ERROR BARS ON BEAM PROPERTIES}

For a fixed error bar, the score is the absolute error $|y - \hat{y}|$ and the error bar is $\hat{y} \pm q$ for every input. For emittance and peak current, we use the error on $\log y$, and the error bar is $\times\!/$\,$c$ (i.e., the simulation is between $\hat{y}/c$ and $\hat{y}c$). We use half of the data for calibration and half for test, over 500 random partitions.

An adaptive error bar changes width with the inputs. We use conformalized quantile regression \cite{romano2019}: two small gradient-boosted tree models predict the 5th and 95th percentiles of the error $y - \hat{y}$ given the 37 inputs, the prediction, and the spread of the 8 sampled beams. The score is how far the error falls outside this range, and the conformal step widens the range by $q$. The bar thus expands and moves where the surrogate is worse. We train on 50\% of the data and calibrate and test on 25\% each, over 40 random partitions. As a baseline, we use the error bar of a model that always predicts the median over all inputs.

\begin{table}[!hbt]
  \centering
  \caption{Conformal error bars at 90\% coverage. Adaptive is the median error bar. Paired values are x and y, or slice and Gaussian fit.}
  \label{tab:bars}
  \small
  \setlength{\tabcolsep}{2pt}
  \begin{tabular}{@{}lcc@{}}
    \toprule
    \textbf{Property} & \textbf{Fixed} & \textbf{Adaptive} \\
    \midrule
    Bunch spacing & $\pm$ 7.0~\textmu m & $\pm$ 6.6~\textmu m \\
    Drive emittance x, y & $\times\!/$\,1.37, 1.44 & $\times\!/$\,1.36, 1.40 \\
    Witness emittance x, y & $\times\!/$\,1.42, 1.39 & $\times\!/$\,1.43, 1.36 \\
    Drive peak current, slice, fit & $\times\!/$\,4.95, 2.55 & $\times\!/$\,2.34, 2.13 \\
    Witness peak current, slice, fit & $\times\!/$\,2.88, 1.53 & $\times\!/$\,2.00, 1.33 \\
    Drive dispersion x, y & $\pm$ 37.0, 4.5~mm & $\pm$ 24.0, 3.7~mm \\
    Witness dispersion x, y & $\pm$ 45.9, 10.1~mm & $\pm$ 39.0, 8.4~mm \\
    \bottomrule
  \end{tabular}
\end{table}

All bars in Table~\ref{tab:bars} have a coverage of 89.5 to 90.9\% on the test data. Compared to the baseline, the error bars are small for the bunch spacing (3\%) and the dispersion (6 to 20\%) (Fig.~\ref{fig:bars}(a)). The method is weakest on the peak current, which the surrogate underestimates: the median drive slice peak prediction is 0.43 of the simulation. The adaptive error bars cut the drive slice peak current error bar from $\times\!/$\,4.95 to $\times\!/$\,2.34 and the dispersion error bars by 15 to 35\%. The bunch spacing and emittance error bars barely change, because their errors do not depend strongly on the inputs.

\subsection{Dependence on Bunch Compression}

The drive bunch errors mainly depend on the L2 phase, and the witness bunch errors on the L1 phase, because these determine the compression. In the fifth of cases with the smallest L2 phase, the drive bunch is long (94~\textmu m rms), and the slice peak current is predicted within 2\%. For L2 phases above $-36.7^\circ$, the bunch is compressed to 35 to 42~\textmu m, and the prediction is only 0.30 to 0.41 of the simulated value (Fig.~\ref{fig:bars}(b)). The compression creates sharp peaks in the simulated current profile, which raise its slice peak current from 1.4 to 3.3 times the Gaussian fit peak. For the surrogate, with the same number of particles, this ratio is only 1.2 to 1.4: it learns the shape of the current profile but smooths out the spikes. Here the fixed prediction intervals fail, since their 90\% coverage is only an average. In the top fifth of L2 phase, they cover the drive dispersion in only 62\% (x) and 73\% (y) of cases, and the adaptive intervals in 80\% and 82\% (Fig.~\ref{fig:bars}(c)).

\section{REGION IN 6D PHASE SPACE}

A prediction interval captures one quantity, but the plasma stage is determined by the whole beam. The invertibility of the flow naturally motivates a region in 6D phase space (Fig.~\ref{fig:region}):
\begin{enumerate}\setlength{\itemsep}{0pt}\setlength{\parsep}{0pt}
  \item For every calibration case, map the simulated particles back through the flow to the Gaussian space at the mean of the stochastic input.
  \item Score the case by the radius of the sphere that contains 95\% of the particles. For a perfect surrogate, the particles would follow a spherical 6D Gaussian, and this radius would be 3.55.
  \item Define the conformal threshold $r$ as the quantile of the scores.
  \item For a test case, map the 6D ball of radius $r$ through the flow to find the region.
\end{enumerate}
If the score of a case is below $r$, at least 95\% of its particles are within the ball, hence within the region. Equation (1) then guarantees that for at least 90\% of test cases, at least 95\% of the simulated electrons are within the region. The region follows the shape of the predicted beam because it is constructed with the same flow. Earlier work constructs a ball in the Gaussian space of a flow for a single vector-valued output \cite{fang2025,dheur2025}. Here the output is a whole bunch, and the score is a property of the bunch.

The thresholds $r$ are 4.45 (drive) and 4.67 (witness). On test cases, the region captures 95\% of the electrons for 89.5\% (drive) and 89.9\% (witness) of beams. For comparison, we construct a Gaussian ellipsoid from the mean and covariance of the predicted beam in the same way, so it has the same guarantee. The median volume of the flow region is 0.19 (drive) and 0.49 (witness) of the ellipsoid volume, since the ellipsoid cannot capture the shape of the beam in the longitudinal phase space (Fig.~\ref{fig:region}(d)). As before, the coverage varies with compression, from 81\% to 98\% across fifths of L2 phase for the drive bunch.

\section{CONCLUSION}

We used conformal prediction to quantify the uncertainty of a normalizing flow surrogate of the FACET-II two-bunch beam. Prediction intervals for beam observables and a region in the full 6D phase space both achieve the specified coverage, and the 6D region is much smaller than a Gaussian ellipsoid with the same guarantee. The prediction intervals need only a model and an independent set of calibration data, so they apply to any accelerator surrogate, and the 6D region applies to any flow surrogate. There are two caveats. First, the bounds hold for the simulator, not the experiment. Second, they are only valid when the data are exchangeable with the calibration data, which is not true for a reinforcement learning controller. Next, we will test the method along the trajectories of such a controller and calibrate it on experimental data.

\section{ACKNOWLEDGMENTS}

The authors thank the Stanford Intelligent Systems Laboratory (SISL) and the FACET-II team at SLAC National Accelerator Laboratory for helpful discussions and feedback. This work was supported in part by the U.S. Department of Energy under Contract No. DE-AC02-76SF00515.

\end{document}